\documentclass[%
reprint,
 amsmath,amssymb,
 aps, physrev,
 pre,
]{revtex4-2}
\usepackage[english,main=english]{babel}
\usepackage{graphicx}
\usepackage{svg}
\usepackage{dcolumn}
\usepackage{bm}
\usepackage{textcomp} 
\usepackage[breaklinks=true]{hyperref}
\usepackage{diagbox}
\usepackage{float}
\usepackage{algorithm}
\usepackage{algpseudocode}

\usepackage{color}
\usepackage{xcolor}

\newcommand{\Reff}{R_{\textrm{eff}}}

\newcommand{\Rtot}{R_{\textrm{tot}}}

\algrenewcommand\algorithmicrequire{\textbf{Input:}}
\algrenewcommand\algorithmicensure{\textbf{Output:}}

\begin{document}

\preprint{APS/123-QED}

\title{\textbf{Effect of Local Topological Changes on  Resistance in Spatially-Embedded Disordered Networks} 
}%

\author{Chenxi Wang}
\author{Charles Emmett Maher}%
\author{Katherine A. Newhall}
 \email{Contact author: knewhall@unc.edu}
\affiliation{Department of Mathematics, University of North Carolina at Chapel Hill, Chapel Hill, NC, USA}

\date{\today}

\begin{abstract}

Disordered materials occur naturally and also provide a broader design space than ordered or crystalline structures. We investigate a two-dimensional disordered network metamaterial constructed from a Delaunay triangulation of an underlying point cloud. Small perturbations in the point cloud induce discrete topological changes. 
One such change we identify is a Delaunay flip, in which two neighboring Delaunay triangles that form a convex quadrilateral structure with their common edge being one of the two quadrilateral diagonals exchange this diagonal for the other diagonal.  
These topological changes can cause substantial jumps in the effective resistance measured diagonally across the network, when the change is located near the source or the sink node. The jumps are explained analytically by showing that the change in effective resistance from edge removal or addition depends on the voltage drop across that edge. However, Delaunay flips have less impact on global resistance measurements and in larger networks. These local topological changes are relevant for finite-sized samples and experimentally-measurable properties such as electrical transport. 
Global characterizations of the network disorder or topology lack the location-specificity of our observed effects on network transport, and thus may be 
inadequate for predicting certain experimentally measurable transport properties in disordered network metamaterials, highlighting the importance of localized regions in material design.

\end{abstract}

\maketitle

\section{Introduction}
Disordered networks are used to model the behaviors of systems across the physical sciences including polymers \cite{broedersz_modeling_2014}, metallic foams \cite{stewart_dynamics_2012}, two-dimensional (2D) materials \cite{moreira_fracturing_2012, obrero_electrical_2025}, actin networks \cite{wang_mechanosensitive_2025}, slime molds \cite{barrows_unifying_2025}, and power grids \cite{nardelli_models_2014} among several others (see, e.g., Ref. \cite{barthelemy_spatial_2022} and references therein for additional examples).
Disorder in network systems has been shown to impart desirable properties including resistance to fracture \cite{fulco_disorder_2025}, exotic vibrational properties \cite{jiao_exotic_2025}, and extraordinary elastic properties \cite{torquato_multifunctional_2018, montiel_effect_2022, skolnick_effective_2025}.
Due to the large design space available to disordered network structures, and their notable physical properties, a number of recent works have focused on characterizing the structure \cite{chieco_quantifying_2021, newby_point_2025, maher_characterizing_2025} and properties \cite{siedentop_stealthy_2024, reid_auxetic_2018, shen_autonomous_2024} of disordered network metamaterials as well as methods to manufacture them \cite{moody_methodology_2025}.

We focus on the electrical transport of network structures.
Such properties have been studied, e.g., through the lens of effective properties for two-phase materials \cite{torquato_multifunctional_2018, skolnick_effective_2025} as well as density functional theory \cite{chen_stonewales_2021}.
Here, we focus on a network science approach, which treats the edges in the network as either wires or lumped circuit elements (in the metric network case) or resistors (in the combinatorial network case).
The metric network framework has been used to model how signals travel through systems like transmission lines and electrical networks \cite{alonso_ruiz_power_2017, chen_power_2017, muranova_effective_2021, muranova_effective_2022, muranova_notion_2020, muranova_eigenvalues_2020, muranova_networks_2022, paul_analysis_nodate, strub_modeling_2019, Bttcher2025}.
Numerous studies have focused on the resistance of combinatorial networks in a mathematical setting (see, e.g., Refs. \cite{barthelemy_spatial_2022,bartkowiak_nonlinear_1995,morris_transport_2012,he_cavd_2020, masuda_random_2017, doyle_random_1984, garcia-redondo_effective_2025}), but fewer studies have used a network framework to explicitly model the electrical transport in materials, e.g., Refs. \cite{obrero_electrical_2025, gu_random-resistor_2022}.
Beyond being interesting in their own right, the same network-based equations that describe electrical transport also describe absorption probabilities in random walks and equilibrium for heat transport.

The method we employ for algorithmically creating disordered network structures is either a Delaunay or Voronoi tessellation of a disordered point cloud.  These tessellations also arise in modeling real materials. 
An example is modeling random composite materials of conducting and nonconducting regions as Voronoi cells \cite{winterfeld_percolation_1981,jerauld_percolation_1984,jerauld_percolation_1984-1}.
To span a range of local disorder, the point cloud can be evolved under Lloyd's algorithm to approach order, as was done in \cite{obrero_electrical_2025}, or perturbations can be made away from a crystalline structure with the size of the perturbation controlling the amount of local disorder.
One example of this kind of point cloud is the uniformly randomized lattice (URL) \cite{klatt_cloaking_2020}, in which uniformly distributed perturbations are applied to each point in a lattice.
The combination of the Delaunay triangulation and evolution of a totally uncorrelated point cloud via repeated applications of Lloyd's algorithm \cite{lloyd_least_1982} (see Sec. \ref{sec:networkgeneration} for details) in 
Ref.~\cite{obrero_electrical_2025} produced abrupt topological changes to the network structure, which motivates the present study.
Standard methods of characterizing network structures that typically indicate sharp changes in network properties, such as those indicating the onset of failure processes in materials \cite{papadopoulos_network_2018}, were not found to capture the observed jumps in resistance.
As a result, the authors of Ref. \cite{obrero_electrical_2025} suggested that the observed effective resistance jumps could be attributed to local topological changes in the networks because such changes have been demonstrated in other pervious works to affect the electrical transport \cite{chen_stonewales_2021,torres-sanchez_analysis_2020}, wave transport \cite{gnutzmann_topological_2013}, fracture mechanisms \cite{berthier_forecasting_2019, sanner_less_2025}, and spectral properties \cite{young_dynamical_2024, doyle_random_1984,tejedor_response_2010} of network structures.
We focus on understanding the effect of localized topological changes in a single finite-sized material that will be designed and manufactured for a specific purpose.
This is in contrast to typical approaches to disorder that consider ensemble-averaged statistics or probability distributions.

We demonstrate that the construction of the Delaunay triangulation from a set of points evolved using Lloyd's algorithm results in discontinuous changes in network topology that include the addition of edges along the boundary of a finite-sized network and concerted deletion and addition of single edges (which we call ``Delaunay flips'').
Our findings that topological changes occurring near the nodes between which the effective resistance $\Reff$ is measured coincide with the large jumps in $\Reff$ observed in Ref.~\cite{obrero_electrical_2025} is perhaps to be expected because these regions also have the largest voltage differences across edges.  
However, it is surprising that the magnitude of the jump due to a single topological change is so large (up to 10\%) compared to
any other changes in $\Reff$ due to the movement of nodes between applications of Lloyd's algorithm.
We support these numerical results with an analytical approximation based on the Sherman--Morrison formula \cite{sherman_adjustment_1950}.

We contrast these Delaunay triangulation results with the corresponding Voronoi tessellations of the same point cloud evolving under Lloyd's algorithm, whose structural changes result in a much smoother evolution of $\Reff$ as the point cloud evolves.
Increasing the network size by increasing the number of points in the point cloud also decreases the size of the jump in $\Reff$.
Moreover, we show that the total effective resistance $\Rtot$, which is a global quantification of the electrical transport properties of a network, is much less sensitive to local topological changes than $\Reff$, and does not effectively capture the effect of topology on the experimentally relevant $\Reff$ measurement.
These findings suggest that global structural descriptors or ensemble averages may obfuscate the structural characteristics that are important to certain physical properties, especially those that are relevant to the characterization of the physical properties of finite systems in experimental contexts.

The rest of this paper is organized as follows.
In Sec.~\ref{sec:methods}, we describe the construction of the networks analyzed in this work, how we characterize their effective resistances, and the methods used to study the effect of network topology on effective resistance.
Then, in Sec.~\ref{sec:results}, we characterize the effect of network topology on effective resistance via computation and present a method to analytically approximate the change in effective resistance due to topological changes.
Finally, in Sec.~\ref{sec:conclusions}, we discuss our findings and offer conclusions and outlook for future studies.

\section{Methods}\label{sec:methods}

\subsection{Network Generation\label{sec:networkgeneration}}

We study the disordered network metamaterials created in Ref.~\cite{obrero_electrical_2025}.
These quasi-2D materials have relatively smooth and systematic evolution toward order, measured by the edge or node entropy. To generate such a sequence of networks, we begin with an initial totally uncorrelated point cloud of $N$ points chosen uniformly in area within a 75mm$\times$75mm bounding box.  These points are then 
evolved within the bounding box with Lloyd's algorithm \cite{lloyd_least_1982}, which has been shown to progress logarithmically towards a crystalline configuration \cite{hong_dynamical_2021, klatt_universal_2019}. This algorithm repeatedly computes the Voronoi tessellation \cite{barthelemy_spatial_2022} within the bounding box and moves each point to the centroid of its Voronoi cell.  A Voronoi tessellation involves finding the polygonal regions of space (Voronoi cells) closer to one point than to any other point in the progenitor point cloud.
After $L$ iterations of the algorithm, the points are connected with either a Delaunay triangulation \cite{barthelemy_spatial_2022} or Voronoi tessellation to form a connected network.  
To generate the Delaunay triangulation, sets of 3 points in the point cloud are connected in a triangle if the circumcircle of those three points does not contain any other points in the pattern. 
This geometric constraint is the \textit{Delaunay criterion}.
Note that one can also generate the Delaunay triangulation by adding an edge between points whose Voronoi cells share an edge.
We then treat the edges and nodes of the polygons from these two tessellation schemes as the edges and nodes of our networks.

\subsection{Electrical Transport}

To understand a network's electrical transport properties, we treat the network edges as ohmic resistors (linear relationship between voltage and current) that connect at the nodes.  
Obrero \textit{et al.} \cite{obrero_electrical_2025} derived a mathematical model of the effective resistance diagonally across the network and validated resistance values computed using this model against experimental measurements of the physical samples 3D printed out of Ti-64-4V.
Next, we summarize this derivation based on applying Kirchhoff's conservation of current law at each node and Kirchhoff's voltage law along each edge to calculate both the effective resistance between any two nodes in the network and the total effective resistance, defined as the sum of the effective resistance over all pairs of nodes.  This derivation closely follows the one found in Ref.~\cite{ghosh_minimizing_2008} and is written in terms of the weighted combinatorial graph Laplacian, an adaptation of the more common $\mathcal{L}_U\vec{V}=r\vec{I}$ in the literature~(for example, see Refs.~\cite{newman_finding_2004,newman_networks_2018}) that uses an unweighted graph Laplacian $\mathcal{L}_U$ and a constant resistance $r$ for all edges.

The resistance of the edge connecting node $i$ to node $j$ is
\[
R_{ij} = \frac{\rho \ell_{ij}}{a} \quad \text{for } i,j=1\dots N, \;i\ne j,
\] 
where $\rho$ is the resistivity of Ti-64-4V ($178\,\mu\Omega\;\mathrm{cm}$),  $a$ is the fixed cross-sectional area of each edge ($a = 0.03$~$\mathrm{cm}^2$) and $\ell_{ij}$ is the length of each edge.
These specific parameter choices correspond to the 3D printed networks from Ref.~\cite{obrero_electrical_2025}. 
By defining the weighted  adjacency matrix as 
\[
A_{ij} = 
\begin{cases}
\frac{1}{R_{ij}} & \text{if } i \text{ and } j \text{ are connected}, \\
0 & \text{otherwise}
\end{cases}
\]
Kirchhoff's laws lead to the equations
\[
\sum_{j=1}^{N} A_{ij} ({V_i - V_j}) = I_i  \quad \text{for } i = 1 \dots N, 
\] 
where $V_{i}$ and $V_{j}$ are the voltage at nodes $i$ and  $j$ respectively, and $I_{i}$ is the current injected into node $i$.
Defining the weighted degree matrix $D$ with diagonal elements equal to the sum of the weights of the edges connected at node $i$,
$D_{ii} = \sum_{j=1}^{N} A_{ij}$ and the weighted graph Laplacian matrix $\mathcal{L}=D - A$, the above in matrix form is
\begin{equation}\label{eq:LVI}
\mathcal{L} \vec{V} = \vec{I},
\end{equation}
where $\vec{V}$ and $\vec{I}$ are the vectors of the voltages and injected currents at each node $i$.  Note that for a network with only one connected component the Laplacian matrix is rank $N-1$ \cite{newman_networks_2018}, consistent with the fact that only the voltage {\em difference} can be uniquely determined; in practice, we set the sink node voltage to zero and solve for the remaining $N-1$ voltages.

\begin{figure}[t!]
\centering
\includegraphics[width=0.45\textwidth]{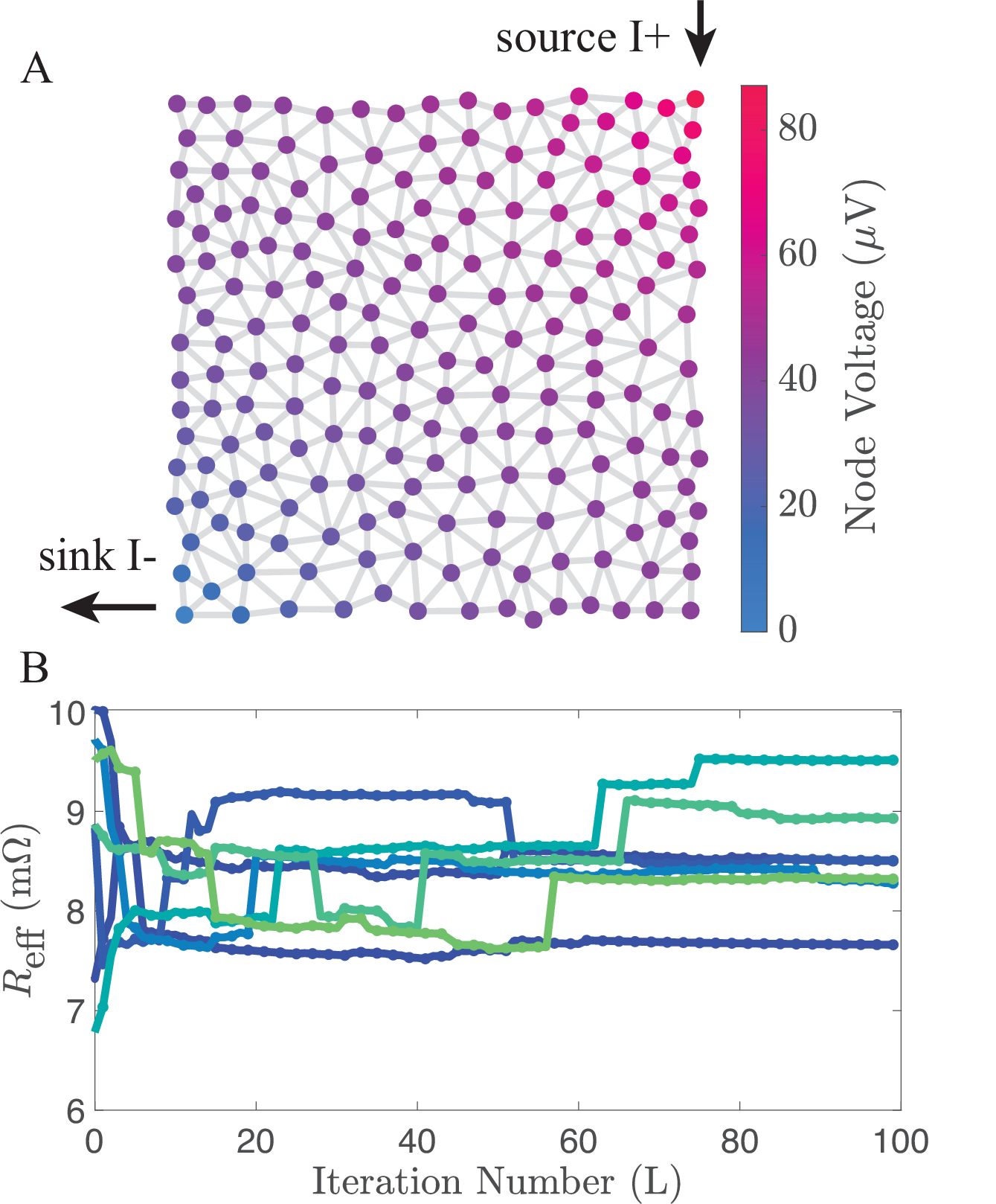}
\caption{A: Schematic showing the source and sink nodes used to compute $\Reff$ diagonally across the network.  Nodes are colored by their voltage, computed with Eq.~\eqref{eq:LVI}.  B: Effective resistance $\Reff$ as a function of iteration number L, showing jumps between successive iterations.  These have been drawn as curves rather than as their underlying discrete points to follow the evolution of each different initial point cloud; they do not represent an interpolation.}
\label{fig:N200Ensemble}
\end{figure}

To compute the effective resistance between any two nodes $i$ and $j$, we inject $I_0$ of current into node $i$, and remove the same amount from node $j$, so that 
the current injection vector $\vec{I}$ is 
\begin{equation}\label{eq:current} 
I_{k} = 
\begin{cases} 
I_{0} &\text {if } k = i,\\ 
- I_{0}&\text {if } k = j,\\
0 &\text {otherwise.}
\end{cases}
\end{equation}
By solving the linear system for $\vec{V}$, we obtain the effective resistance between any two nodes $i$ and node $j$ as
\[\Reff^{ij} = \frac{|V_{i} -V_{j}|}{I_{0}}. \] 
While $\Reff$ is independent of the choice of $I_0$, a value needs to be chosen to set up the linear system in Eq.~\eqref{eq:LVI} and solve it numerically; we use $I_0=10$ mA to be consistent with prior work in 
Ref.~\cite{obrero_electrical_2025}.
Additionally, we are specifically interested in the experimental configuration A from Ref.~\cite{obrero_electrical_2025}, thus we refer to $\Reff$ as the effective resistance diagonally across the network from the northeast-most node to the southwest-most node.  One example of this configuration, and the resulting voltage at each node, is shown in Fig.~\ref{fig:N200Ensemble}A.

We also calculate the total effective resistance, also called the Kirchhoff's index,  that quantifies the network's global electrical properties. 
It is the pairwise sum of all the effective resistances, 
\[
R_{\text{tot}} = 
\sum_{1 \leq i < j \leq N} \Reff^{ij},
\] 
which is equivalent to $N$ times the sum of the inverse of the non-zero eigenvalues $\lambda_k$ of the weighted graph Laplacian matrix~\cite{klein_resistance_1993,gutman_quasi-wiener_1996},
\begin{equation}\label{eq:Rtot}
R_{\text{tot}}  = N \sum_{k=2}^{N} \frac{1}{\lambda_k}.
\end{equation} 
While a number of idealizations are made in the construction of this model (e.g., no contact resistance at nodes and no capacitive effects), we note that previous work has shown that the model has good agreement with experimental data \cite{obrero_electrical_2025} and thus our model can predict the effective conductivity of network-based materials.

\subsection{Edge Flip Enumeration Algorithm\label{sec:edgefliptest}}

Here we describe the {\em edge flip enumeration test}, which we designed to show the spatially heterogeneous response of single topological changes on either $\Reff$ or $\Rtot$.  
Algorithm 1 first enumerates all network edges that are part of two neighboring triangles that form a convex quadrilateral with their common edge being one of the two quadrilateral diagonals.
Next, each enumerated diagonal edge is independently flipped to the alternate diagonal, and the resulting resistance changes are recorded.
This topological change is similar to the Delaunay flip we observe between successive iterations of Lloyd's algorithm; however, the edge flip enumeration test can be performed on any network with the aforementioned convex quadrilateral structures.
More generally, one could probe the effect of removing any edge in the network and adding another edge as long as the added edge does not cross an existing edge, thus keeping the network spatially embedded.  Note that since we manually impose the diagonal flip, if applied to a Delaunay triangulation, the resulting network is now in violation of the Delaunay criterion.

  \bigskip
\noindent\rule{\linewidth}{0.2pt}
{\bf Algorithm 1:} Edge Flip Enumeration Test
\noindent\rule{\linewidth}{0.2pt}
\begin{algorithmic}[1]
  \Require  $A$, the weighted adjacency matrix of the network
  \Ensure  ${\Delta \Reff, \Delta \Rtot}$ resistance change for each diagonal edge flip
  \State Compute original $\Reff$, $\Rtot$
  \State Enumerate all triangles in the network
  \State Enumerate all edges connecting nodes $i$ and $j$ with ($i < j$)
  \State Form a mapping between each edge and the triangles the edge belongs to.
  \For{each edge $e_{ij}$}
    \If{$e_{ij}$ belongs to at least two triangles}
      \State Check convexity of the quadrilateral 
      \If{convex}
        \State Delete edge $e_{ij}$
        \State Add edge of opposite diagonal in  quadrilateral
        \State Compute new $\Reff$, $\Rtot$; calculate $\Delta \Reff, \Delta \Rtot$
        \State Return network to its original state
      \EndIf
    \EndIf
  \EndFor
\end{algorithmic}
%
\noindent\rule{\linewidth}{0.2pt}

\subsection{Tortuosity\label{sec:tortuosity}}

Tortuosity is the ratio of the shortest-path distance between two nodes in a spatially embedded network and the Euclidean distance between the two nodes.  Therefore, a tortuosity value of one indicates a perfectly straight path between two network nodes, while a larger value corresponds to a more indirect path.
We calculate the change in tortuosity across a convex quadrilateral structure in a network when the edge flip test is applied to that quadrilateral as a way to help explain the magnitude of effective resistance change.  Specifically, we identify quadrilaterals whose 
sets of nodes contain either the source or sink node  for a network.  
Then, we label the quadrilateral nodes $a$-$c$-$b$-$d$ with node $a$ the source or sink node, depending on the location in the network of the identified quadrilateral.  If there is an edge connecting node $a$ to node $b$, the tortuosity is equal to one, as the network path is the same as the Euclidean length.  If there is not an edge, then the tortuosity is
 \[
\tau_{ab} =  \frac{C_{ab}}{l_{ab}},
\]
where $l_{ab}$ is the Euclidean distance between nodes $a$ and $b$ and
the path length along the quadrilateral edges is
\[
C_{ab} = \min\Bigl(
l_{ac} +l_{bc}, l_{ad}+l_{bd}
\Bigr).
\] 
Thus, the change in tortuosity when the edge flip test is applied to the quadrilateral is
\begin{equation}\label{eq:delta_tau}
\Delta \tau_{ab} = | 1 - \tau_{ab} |.
\end{equation}

\section{Results}\label{sec:results}

\subsection{Jumps in Effective Resistance are caused by local topological changes near source/sink nodes\label{sec:resultsA}}

Using Lloyd's algorithm and the Delaunay triangulation to create a sequence of networks that begin disordered and progress towards ordered, Obrero \textit{et al.} \cite{obrero_electrical_2025} observed large jumps in the effective resistance between sequential Lloyd's iterations.  These jumps can be seen in Fig.~\ref{fig:N200Ensemble}B, where effective resistance, $\Reff$, for an $N=200$ node network is plotted vs.~Lloyd's iteration number, $L$, and each curve corresponds to a different initial point cloud.  
Note that these have been drawn as curves rather than as their underlying discrete points and should not be interpreted as interpolations between successive Lloyd's iterations.  This applies to all subsequent figures.
While the jumps are accompanied by a topological change in the adjacency matrix (adding or removing edges), not all topological changes in the adjacency matrix correspond to a large jump in $\Reff$.  Here, we explain these jumps, finding that edges near the northeast and southwest corners of the network, where current is applied, have a larger impact on $\Reff$. 
We specifically study these behaviors in individual configurations as opposed to examining ensemble-averaged behaviors because we are interested in understanding how to tailor the electrical transport of individual manufactured samples.
Such a study of the structures of individual ensemble members (the ``geometric structure approach'') has been leveraged elsewhere to, e.g., characterize disordered and mechanically rigid assemblies of nonoverlapping particles (see Ref.~\cite{torquato_physics_2026} and references therein).
Nonetheless, we more carefully examine the ensemble-averaged behavior of our networks in Sec. \ref{sec:finitesize}.

\begin{figure}[b]
\centering
\includegraphics[width=0.45\textwidth]{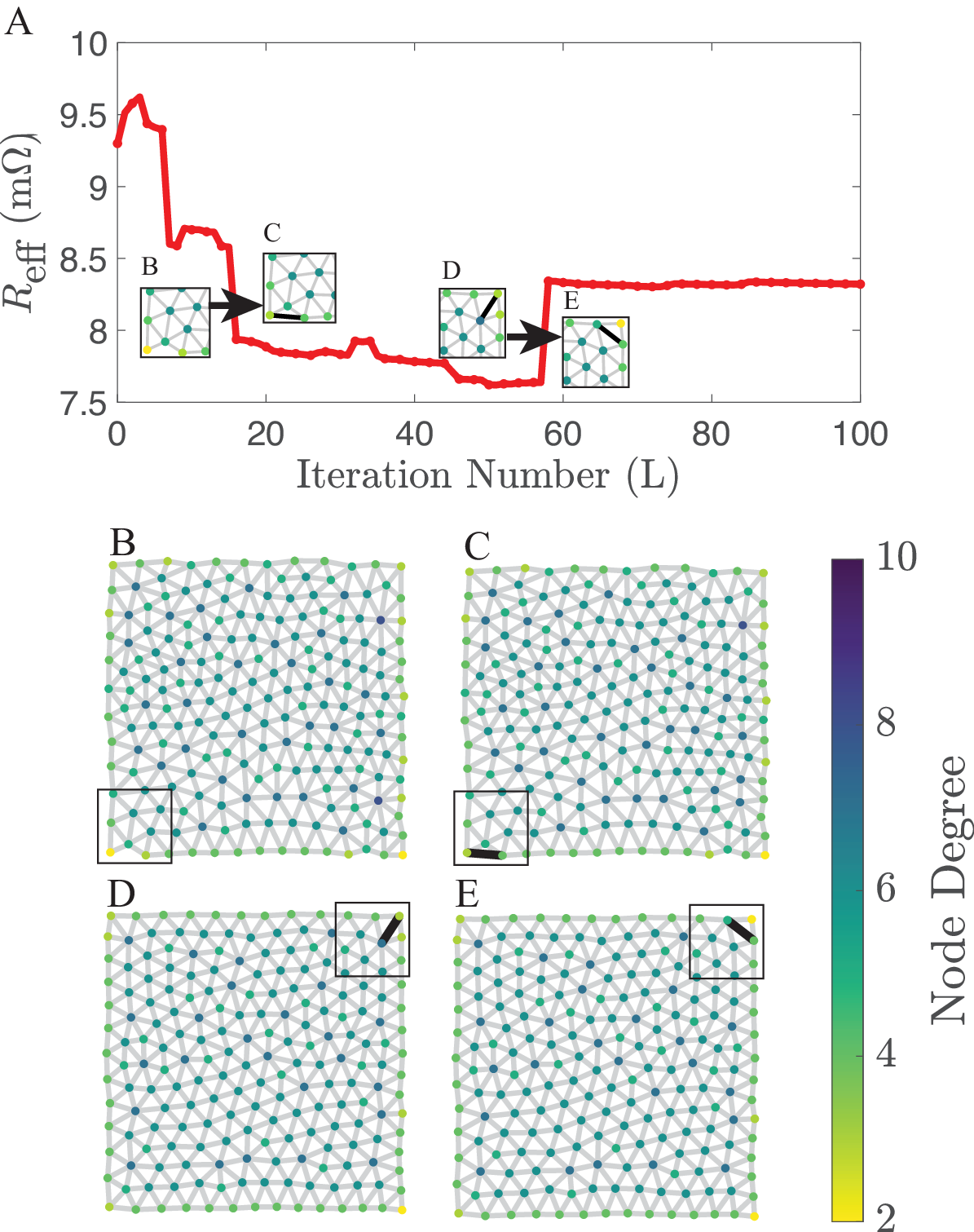}
\caption{A: The progression of $\Reff$ as a function of iteration number L for one simulation, highlighting two jumps and two topological changes.  The insets B and C highlight in black the addition of an edge at the bottom left corner of the network; the full networks are shown in panels B and C.  The insets D and E highlight in black the Delaunay flip in the top right corner of the network; the full networks are shown in panels D and E.}
\label{fig:TopoChangeScheme}
\end{figure}

In Fig.~\ref{fig:TopoChangeScheme}, we show examples of the two topological changes responsible for the large jumps. A third mechanism could be that the source or sink node changes, but we do not explicitly consider this case as it is not a local topological change. Figure~\ref{fig:TopoChangeScheme}A highlights just one initial point cloud (simulation 14 from Ref.~\cite{obrero_electrical_2025}) and two jumps in $\Reff$, one from $L=14$ to $L=15$ and one from $L=57$ to $L=58$.  The networks before and after the first (second) jump are depicted in Figs.~\ref{fig:TopoChangeScheme}B and C (D and E).  The jump labeled B$\to$C coincides with the addition of an edge along the boundary at the southwest corner.  The new edge is marked with a thick black line in Fig.~\ref{fig:TopoChangeScheme}C.  The jump labeled D$\to$E coincides with the flip of one edge inside the quadrilateral structure at the northeast corner.  This flipped edge is marked with thick black lines in Figs.~\ref{fig:TopoChangeScheme}D and E.

As all of the nodes move between successive Lloyd's iterations, we also check how the above-identified topological changes affect $\Reff$ in isolation.  Keeping the network fixed in the configuration before the jump, we manually perform the single topological change (adding a single edge or flipping a single edge) and recompute $\Reff$.  These values of $\Reff$ are summarized in Table~\ref{tab:cornerlabel} for a number of example jumps. 
We hand-select these examples to illustrate some of the largest jumps we observe. 
Other individual topological changes that result in relatively small perturbations to $\Reff$ are difficult to distinguish from changes to $\Reff$ due to the movement of nodes, and thus are not included in Table~\ref{tab:cornerlabel}.
We compute how much of the true jump this single topological change accounts for by dividing the true jump in $\Reff$ by the manual jump.  In all cases, nearly 100\% of the jump is accounted for with this single topological change.  (Accounting for over 100\% simply indicates the topological change in isolation results in a larger jump than observed.) These results provide strong evidence that a single topological change is sufficient to explain the observed large jump in effective resistance, but not all topological changes result in such jumps. This observation motivates our closer examination of the topological mechanism and when it results in large jumps.

\begin{table}[b!]
\centering
\begin{tabular}{|c|c|c|c|c|c|}
\hline
Simulation Index & Type & Before & After & Manual& Accounted\\
of Jump from \cite{obrero_electrical_2025} & ~ &  ($m\Omega$) &  ($m\Omega$) &  ($m\Omega$)& ~\\
\hline
sim3	iter3-4	&	Add	&	9.7011	&	8.7956	&	8.8533	&	93.6\%	\\
sim4	iter6-7	&	Flip	&	8.6439	&	7.8119	&	7.8641	&	93.7\%	\\
sim6	iter9-10	&	Flip	&	7.6916	&	8.3276	&	8.3347	&	101.1\%	\\
sim6	iter12-13	&	Flip	&	8.3214	&	8.974	&	8.9648	&	98.6\%	\\
sim6	iter52-53	&	Add	&	9.0918	&	8.5121	&	8.5127	&	99.9\%	\\
sim7	iter4-5	&	Add	&	8.5839	&	7.8772	&	7.8814	&	99.4\%	\\
sim7	iter20-21	&	Flip	&	7.7698	&	8.5498	&	8.543	&	99.1\%	\\
sim8	iter23-24	&	Flip	&	7.9019	&	8.6116	&	8.5755	&	94.9\%	\\
sim10	iter28-29	&	Flip	&	8.5531	&	7.9436	&	7.9582	&	97.6\%	\\
sim10	iter41-42	&	Flip	&	7.8338	&	8.5768	&	8.5827	&	100.8\%	\\
sim10	iter66-67	&	Flip	&	8.5041	&	9.1118	&	9.1104	&	99.8\%	\\
sim14	iter6-7	&	Add	&	9.3958	&	8.5991	&	8.5937	&	100.7\%	\\
sim14	iter14-15	&	Add	&	8.5796	&	7.9373	&	7.9275	&	101.5\%	\\
sim14	iter57-58	&	Flip	&	7.6367	&	8.3427	&	8.3447	&	100.3\%	\\
\hline
\end{tabular}
\caption{Values of $\Reff$ before and after a topological change in the networks from Ref. \cite{obrero_electrical_2025}.  ``Type'' indicates whether this topological change was the addition of an edge (``Add'') or a Delaunay flip (``Flip''). 
 The columns ``Before" and ``After" contain the true value of $\Reff$ before and after the jump, respectively. The column ``Manual" contains the value after the single topological change identified in column ``Type'' is executed manually. The column ``Accounted" is calculated as 
$(\text{Manual} - \text{Before})/(\text{After} - \text{Before})$.}
\label{tab:cornerlabel}
\end{table}

\begin{figure}[t]
\centering
\includegraphics[width=0.45\textwidth]{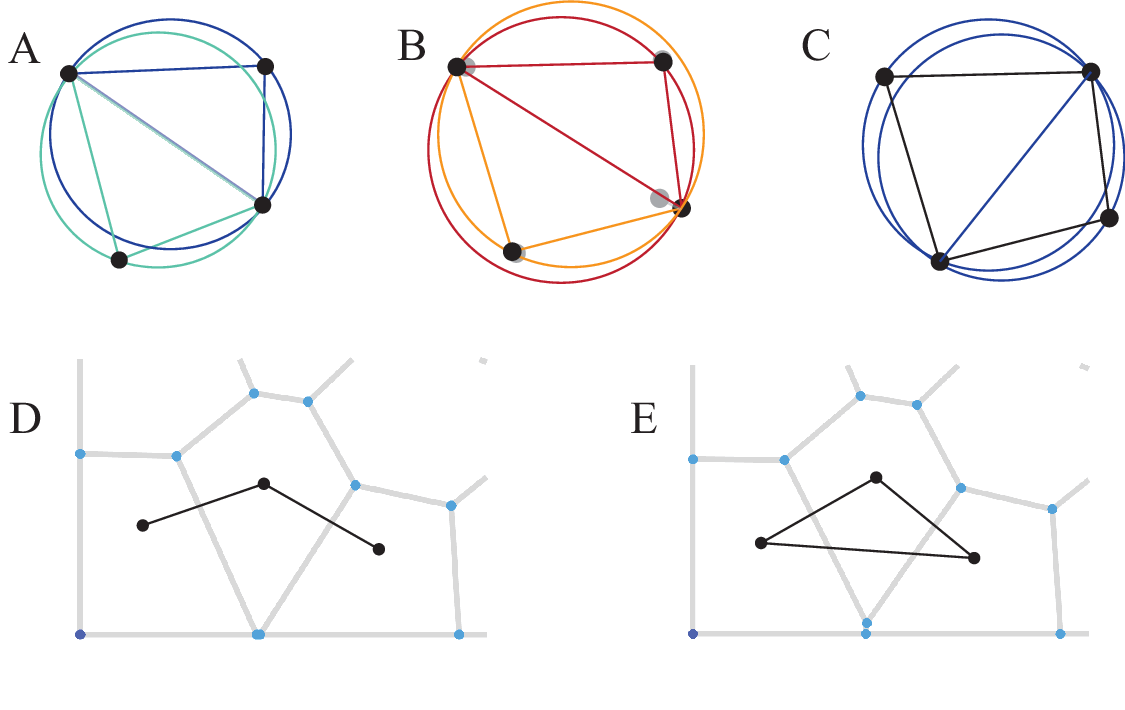}
\caption{Mechanisms of topological changes. Delaunay flip occurs between panel A and C, with panel B illustrating how the Delaunay criterion is violated as a result of one node moving from the gray location.  An edge addition at the boundary occurs between panels D and E, where the Voronoi tesselation is shown with blue nodes and gray edges and the Delaunay triangulation for these three complete cells is shown in black.}
\label{fig:Mechanism}
\end{figure}

The two topological changes (Delaunay flip and edge addition) occur abruptly even though the underlying points of the Delaunay triangulation undergo small changes, especially at higher values of $L$. Figure~\ref{fig:Mechanism} depicts the two mechanisms: panels A, B and C depict the edge flip, a result of the Delaunay criterion as points get perturbed, and panels D and E depict the edge addition, primarily driven by the imposed boundary. 
Figure \ref{fig:Mechanism}A depicts two circumcircles for four points in the form of a quadrilateral with one diagonal. Figure \ref{fig:Mechanism}B shows the same circumcircles but now with the points perturbed from their original gray position to their new black position.  The dark blue (light blue) circle evolves to the dark red (light orange) circle and now contains the lower-left (upper-right) point within it; the Delaunay criterion is not met. 
Flipping the diagonal, shown in Fig.~\ref{fig:Mechanism}C, yields an edge configuration satisfying the Delaunay criterion.
The edge addition is a result of computing the Voronoi tessellation in a bounded, rather than infinite, domain.
Recall connecting points of bordering Voronoi cells is an alternative method from the Delaunay criterion for constructing the Delaunay triangulation.  Figure \ref{fig:Mechanism}D shows part of the Delaunay triangulation in black for the three Voronoi cells at the boundary of the domain.  After the points are perturbed under one iteration of Lloyd's algorithm, the center Voronoi cell no longer intersects the boundary, allowing the two other Voronoi cells to share a boundary. This results in the addition of a third Delaunay edge, as shown in Fig.~\ref{fig:Mechanism}E.

\begin{figure}[t!]
\centering
\includegraphics[width=0.48\textwidth]{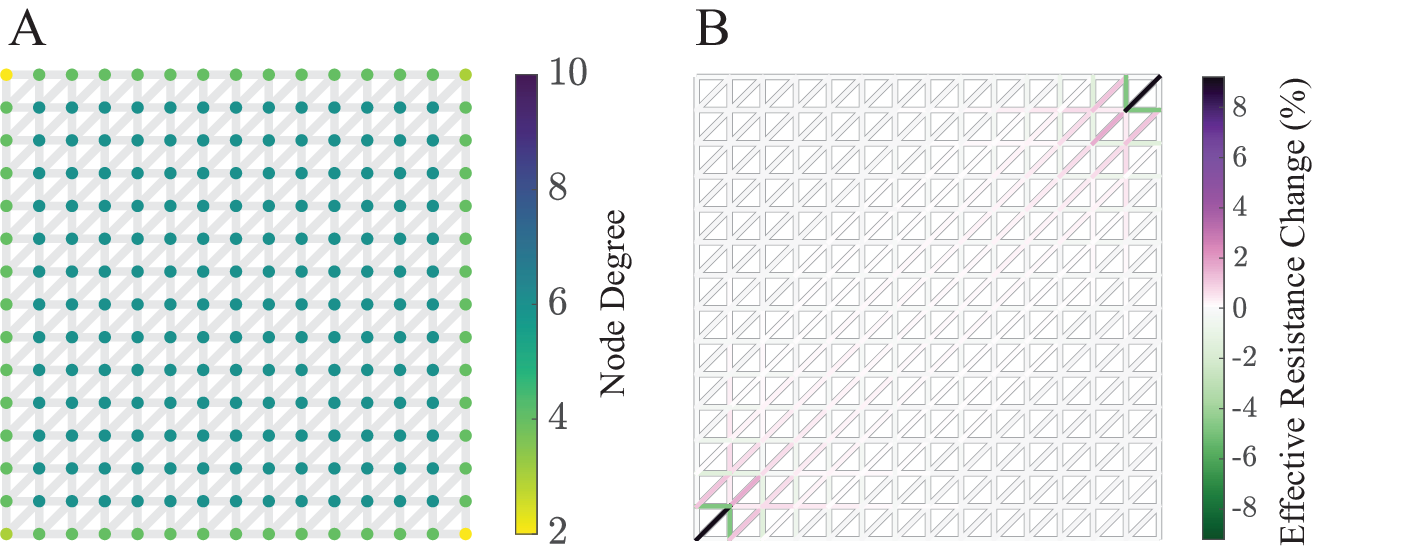}
\caption{A: Simplified lattice network.  B: Heatmap where the color of each edge indicates the relative percent change of the effective resistance, $\Reff$, when the edge is flipped to the other diagonal of its quadrilateral.  Boundary edges are not flipped, and are colored white (i.e. zero change).  
}
\label{fig:LatticeHeatMap}
\end{figure}

Having numerically verified that a single edge flip can account for nearly the entire jump in $\Reff$, we further 
hypothesize that only changes local to the source or sink node cause these large jumps in resistance.  To numerically support this hypothesis, we start with a simpler case: a $15\times 15$ square lattice with the southwest to northeast diagonal of each unit connected, as depicted in Fig.~\ref{fig:LatticeHeatMap}A.  We independently flip each interior edge while keeping the others unflipped, reporting the  percentage change in $\Reff$ in the heat map shown in Fig.~\ref{fig:LatticeHeatMap}B.
We observe that edges along the northeast-southwest diagonal have the largest effect on $\Reff$, with the most significant change being from the edges  directly in contact with the source or the sink node.

\begin{figure}[t!]
\centering
\includegraphics[width=0.48\textwidth]{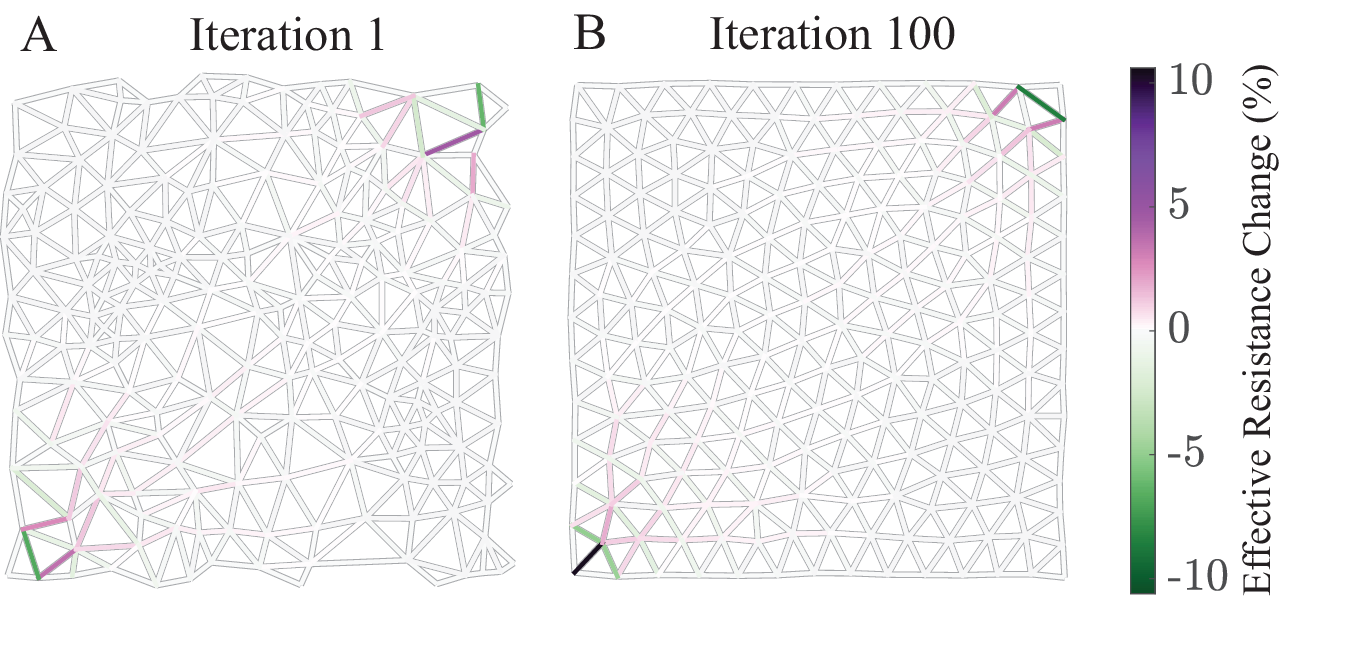}
\caption{Heatmap where the color of each edge indicates the relative percent change of the effective resistance, $\Reff$, when the edge is flipped to the other diagonal of its quadrilateral.  Boundary edges and those not part of a convex quadrilateral are not flipped, and are colored white (i.e. zero change).
}
\label{fig:DisorderedHeatMap}
\end{figure}

To further test this sensitivity to topological perturbation near the source or sink node, we perform a similar flip-test on the random networks generated by the method of Sec.~\ref{sec:networkgeneration}.
First, we identify all convex quadrilaterals in the network as explained in Sec.~\ref{sec:edgefliptest}.  Then, as with the simple lattice network, we independently  flip the diagonal in each quadrilateral while keeping the others unflipped, reporting the percent change in $\Reff$ in Fig.~\ref{fig:DisorderedHeatMap}. 
Consistent with the lattice network, edges near the source and sink nodes have the greatest effect.  Note that in the lattice network, flipping the diagonal of each square cell always resulted in an 
increase in $\Reff$ since each edge was being flipped from parallel to perpendicular to the diagonal of the current application.  In the random networks, we see both increase and decrease depending on edge-orientation; we only present the magnitude of the change. 
This is because in the lattice network, the cell diagonals are all aligned in the direction from the source to the sink node.  Flipping any one diagonal disrupts this and thus increases the effective resistance.  In the disordered networks, the diagonals are not all aligned in the same direction, thus a flip may help or hinder the flow of current from source to sink node.

\subsection{Predicting Location of Topological Change Producing Largest Effective Resistance Jumps\label{sec:predict}}

In Sec.~\ref{sec:resultsA}, we identified that edges near the source and sink nodes have the greatest effect on the effective resistance. Here, we further explore this finding over an ensemble of networks, revealing that closeness to source/sink nodes alone is not enough to predict the magnitude of the effect of a topological change on the effective resistance.  
We analytically approximate the effect of a Delaunay flip on the effective resistance, finding that it depends on the voltage difference across the flipped edge, thereby supporting that the location of edges responsible for the largest changes in $\Reff$ are located near the source and sink nodes where voltage differences are highest, but also that longer edges produce larger differences in effective resistance.

\begin{figure}[t!]
\centering
\hspace*{-0.04\textwidth}
\includegraphics[width=0.48\textwidth]{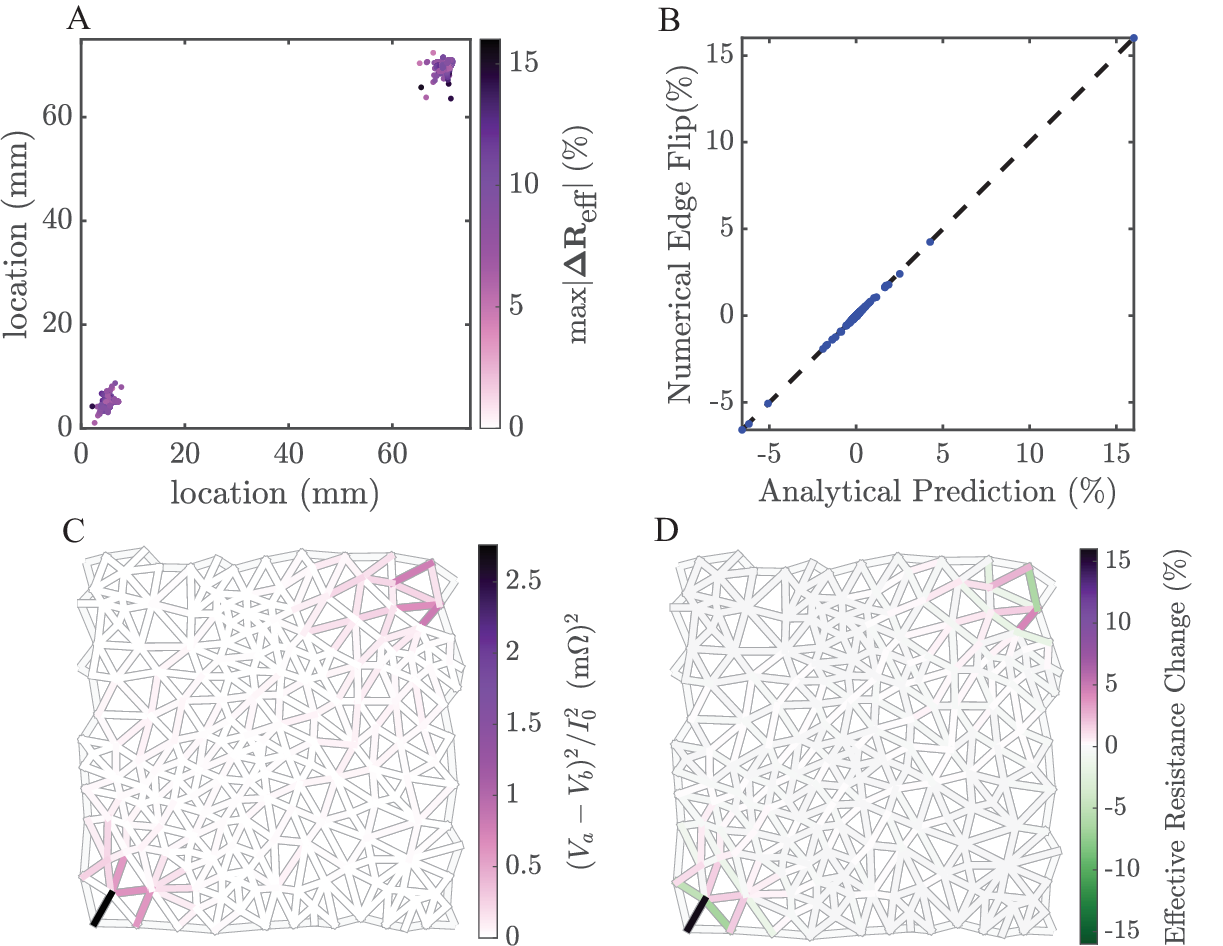}
\caption{A: Scatter plot of the location and the magnitude of the maximum effective resistance change found in each network, under the edge flip enumeration test. These networks are the same as those used in \cite{obrero_electrical_2025}. B: 20 different initial point clouds at each iteration number $L=0$ to $L=100$.
B: Scatter plot comparing the analytical prediction in Eq.~\eqref{eq:twoDeltaR} to the numerical edge flip of effective resistance change for each edge in the network shown in panels C and D. C: Heatmap of the voltage difference scaled by the applied current, $(V_a-V_b )^2 /I_0^2$ , for each edge in the network.
D: Heatmap of Eq.~\eqref{eq:twoDeltaR} scaled by the initial effective resistance.  
Boundary edges and those not part of a convex quadrilateral are colored white (i.e. zero) since there is no direct comparison to the edge flip test. 
}
\label{fig:Quantify}
\end{figure}

In Fig.~\ref{fig:Quantify}A, we plot the location of the center of the edge responsible for the maximal magnitude of change in $\Reff$ when performing the edge flip test.  This was performed on 20 simulations at each iteration of Lloyd's algorithm, $L=0$ to $L=100$.
All edges responsible for maximal change are in the vicinity of the source or sink node. 
To further understand and predict these network features that affect drastic changes on the effective resistance, we calculate the effective resistance change in terms of the weighted graph Laplacian of the network.

To analytically predict the effect of a topological change, we start by adapting formulas for the effective resistance from Refs.~\cite{klein_resistance_1993,ghosh_minimizing_2008}. 
Recall that the weighted Laplacian $\mathcal{L}$ is rank $N-1$.  We use a tilde in this section to indicate that one node has voltage set to zero (grounded), and its corresponding element is removed from all vectors and matrices.  For example, if node $j$ is grounded then
$$
\tilde{\mathcal{L}}_{i,k} = \left\{ \begin{array}{ll} \mathcal{L}_{i,k} & \textrm{if } i,k < j, \\
\mathcal{L}_{i-1,k} & \textrm{if } i>j, k < j, \\
\mathcal{L}_{i,k-1} & \textrm{if } i<j, k > j, \\
\mathcal{L}_{i-1,k-1} & \textrm{if } i,k > j .
\end{array}
\right.
$$
While the effective resistance does not depend on the choice of applied current $I_0$, we include it in this derivation to keep units consistent.  We write the current injection vector from Eq.~\eqref{eq:current} as $\vec I = I_0(\tilde{\mathbf e}_i -\tilde{\mathbf e}_j)$
where $\mathbf{e}_i$ is the standard basis vector with $1$ in element $i$ and zero elsewhere.  Then the effective resistance across node $i$ and $j$ is 
\[
R_{\mathrm{eff}}^{i j} = \frac{(\tilde{\mathbf e}_i -\tilde{\mathbf e}_j)^\top\tilde{\mathcal{L}}^{-1} I_0 (\tilde{\mathbf e}_i -\tilde{\mathbf e}_j)}{I_0}.
\]
Taking into account that node $j$ is grounded ($\tilde{\mathbf e}_j$ is a vector of zeros) we have that
\[
R_{\mathrm{eff}}^{i j}  = \left[ \tilde{\mathcal{L}}^{-1} \right]_{ii}.
\]
An edge flip first deletes an edge, then adds an edge.  Deleting the edge connecting nodes $a$ and $b$ is a rank-1 update to the weighted graph Laplacian, given by
\[
\tilde{\mathcal{L}'} = \tilde{\mathcal{L}} - \frac{1}{R_{ab}}  (\tilde{\mathbf{e}}_a - \tilde{\mathbf{e}}_b)(\tilde{\mathbf{e}}_a - \tilde{\mathbf{e}}_b)^\top,
\]
where $R_{ab}$ is the true resistance of edge $ab$.
Then, using the Sherman--Morrison formula \cite{sherman_adjustment_1950}, we rewrite the inverse of this rank-1 updated matrix as the inverse of the original matrix with a rank-1 update, given by 
\[
{\mathcal{L}'}^{-1} = \left[ \tilde{\mathcal{L}}^{-1} \right]_{ii}
+ \frac{\frac{1}{R_{ab}}  \left( \left[ \tilde{\mathcal{L}}^{-1} (\tilde{\mathbf{e}}_a - \tilde{\mathbf{e}}_b) \right]_i \right)^2}{1 - \frac{1}{R_{ab}}  (\tilde{\mathbf{e}}_a - \tilde{\mathbf{e}}_b)^\top \tilde{\mathcal{L}}^{-1} (\tilde{\mathbf{e}}_a - \tilde{\mathbf{e}}_b)}.
\]
Note that 
\[
(\tilde{\mathbf{e}}_a - \tilde{\mathbf{e}}_b)^\top \tilde{\mathcal{L}}^{-1} (\tilde{\mathbf{e}}_a - \tilde{\mathbf{e}}_b) \equiv \Reff^{ab}
\]
and
\[
\left[ \tilde{\mathcal{L}}^{-1} (\tilde{\mathbf{e}_a} - \tilde{\mathbf{e}_b}) \right]_i  \equiv \frac{V_a - V_b}{I_0}.
\]
Thus, $\Delta R_{\mathrm{eff}}^{i j} = \left[ \tilde{\mathcal{L}}'^{-1} \right]_{ii} - \left[ \tilde{\mathcal{L}}^{-1} \right]_{ii}$ is given by
\begin{equation}\label{eq:oneDeltaR}
\Delta R_{\mathrm{eff}}^{i j} 
 = \frac{(V_a - V_b)^2}{I_0^2}\frac{1}{R_{ab} - R_{\mathrm{eff}}^{a b}}.
\end{equation}
An edge flip is an edge deletion followed by an edge addition, so we can approximate the change in $\Reff$ under an edge flip from edge $ab$ to edge $cd$ by 
\begin{equation}\label{eq:twoDeltaR}
\Delta R_{\mathrm{eff}}^{i j} \approx \frac{(V_a - V_b)^2}{I_0^2(R_{ab} - R_{\mathrm{eff}}^{a b})}-\frac{(V_c - V_d)^2}{I_0^2(R_{cd} +R_{\mathrm{eff}}^{c d})}.
\end{equation}
 
Equation~\eqref{eq:twoDeltaR} fully accounts for the change in effective resistance, as shown in Fig.~\ref{fig:Quantify}B that compares to the result of the edge flip test.
Note that Eq.~\eqref{eq:twoDeltaR} would be exact if $\Reff^{cd}$ were to be recomputed after the deletion of edge $ab$.
From Eq.~\eqref{eq:twoDeltaR} we also see that a simple physical quantity---the edge voltage difference---is a strong predictor of the magnitude of $\Delta \Reff$.
This explains why edges near the source and sink nodes have the largest 
effect, as they too have the largest voltage difference.  This is confirmed in Fig.~\ref{fig:Quantify}C, where we plot $(V_a - V_b)^2/I_0^2$, the square of the voltage difference across each edge scaled by 
$I_0$.  Note, edges in Fig.~\ref{fig:Quantify}D that have a negative change on $\Reff$ are missing in Fig.~\ref{fig:Quantify}C since the behavior captured by the second term in Eq.~\eqref{eq:twoDeltaR}---the voltage drop across the edge in its flipped orientation---is not accounted for.

By further investigating individual networks with some of the largest maximum effective resistance changes, we
 notice that edge length, which is proportional to resistance, also plays a role in the magnitude of the change in effective resistance. 
Specifically, networks with high maximum percentage change seem to have highly oblique quadrilaterals with one short and one large diagonal.  One such example appears in the lower-left corner of the network in Fig.~\ref{fig:Quantify}D.

\subsection{Extensions to Different Scenarios}
The edge flip test and Eq.~\eqref{eq:twoDeltaR} can easily be extended to other scenarios.  In this section, we consider two such scenarios, one in which the effective resistance is measured between an interior and a corner node.  Another, in which the point cloud is generated by perturbing a triangular lattice rather than using Lloyd's algorithm.
These two extensions can already be treated with our analytical framework.
Equation \eqref{eq:twoDeltaR} is already written generically to handle the effective resistance between any pair of nodes $i$ and $j$.
The network structure enters into Eq.~\eqref{eq:twoDeltaR} through the graph Laplacian $\mathcal{L}$.  Having already shown the agreement of Eq.~\eqref{eq:twoDeltaR} and the edge flip test, we only present results for the edge flip test next.

\begin{figure}[t]
\centering
\includegraphics[width=0.5\textwidth]{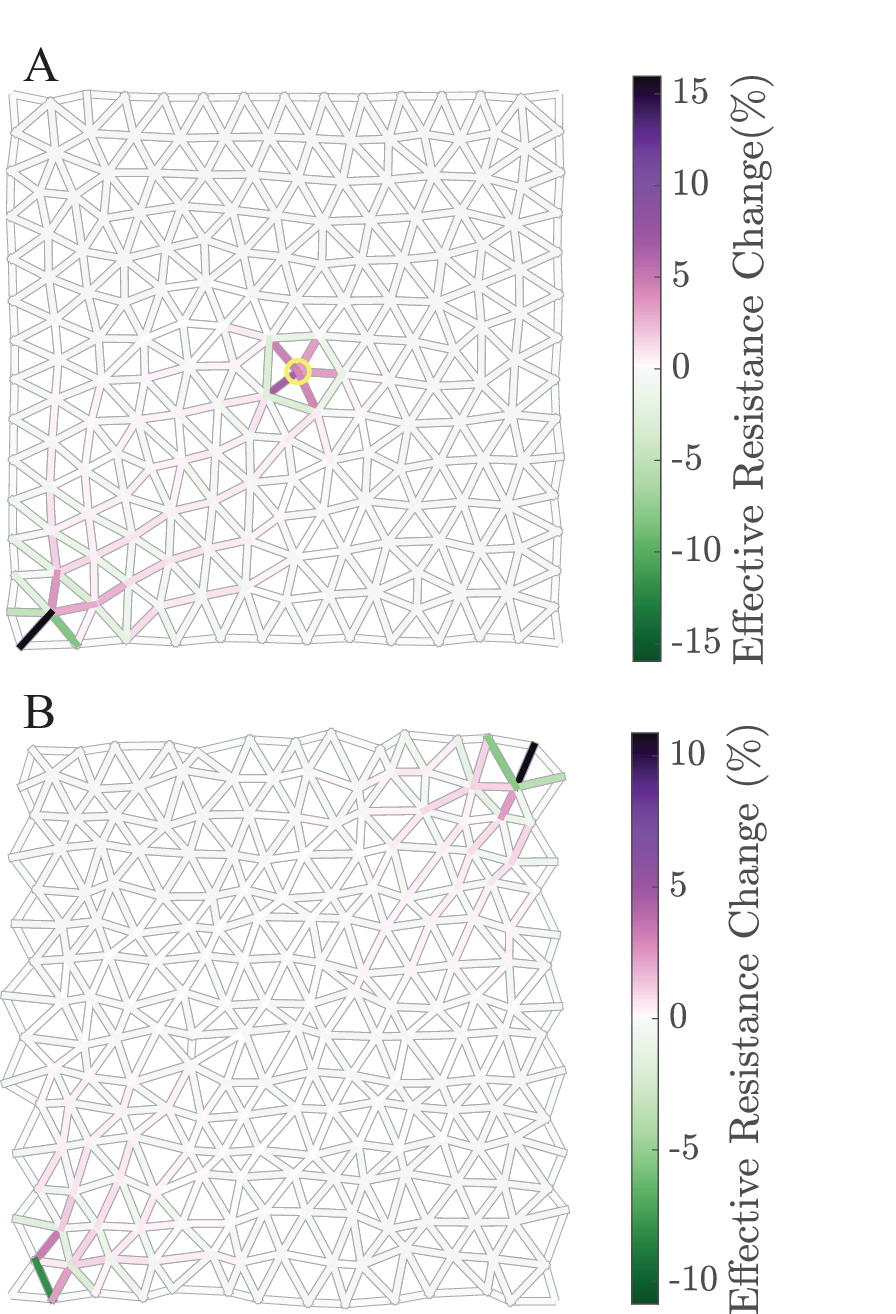}
\caption{A: For the case of effective resistance measured between the center node circled in yellow and the southwest corner node, the heatmap indicates the relative percent change of this effective resistance. B: Heatmap of URL network where the color of each edge indicates the relative percent change of effective resistance $\Reff$, when the edge is flipped to the other diagonal of the quadrilateral. Boundary edges and those not part of a convex quadrilateral are not flipped, and are colored white (i.e. zero change). }
\label{fig:extensions}
\end{figure}

The edge flip algorithm can be extended to track changes to the effective resistance measured between any pair of nodes, simply by changing the source and sink nodes.  This is done by changing $i$ and $j$ in Eq.~\eqref{eq:current}.  Figure \ref{fig:extensions}A shows one such example where the edge flip test has been applied when measuring the effective resistance between the circled center node and the southwest corner node. 
Despite using an interior node as the source node, we still see the same effect that edges near the source and sink nodes result in the largest change to effective resistance. Moreover, the effect is stronger for the edges between the source and sink nodes. Since the effect remains unchanged when moving the source node away from the boundary, it is reasonable to suspect that under periodic boundary conditions the effect would also remain unchanged. 
In this work, we are primarily concerned with physically manufacturable network materials; therefore, we do not directly implement a test of this effect under periodic boundary conditions.

The edge flip algorithm is also not limited by how the disordered network is generated.  As another example, we generate a disordered point cloud by starting with $N=216$ points in a triangular lattice within the bounding box.  Then each point is perturbed to a uniformly random location within a small box with side length 2.85mm centered at the lattice point before connecting the points with a Delaunay triangulation.
The result of the edge flip test on this network is shown in Fig.~\ref{fig:extensions}B.  We see behaviors consistent with Figs.~\ref{fig:LatticeHeatMap}B, \ref{fig:DisorderedHeatMap}, ~\ref{fig:Quantify}D, and \ref{fig:extensions}A, suggesting that the importance of the edges in the vicinity of the source and sink nodes is agnostic to the way disorder is introduced to the point pattern.  Note that Voronoi tessellations, that we consider in Sec.~\ref{sec:voronoi}, are outside the scope of the edge flip test since they generally do not contain the required cell geometries.

\subsection{Finite-Size Effects}\label{sec:finitesize}

We have shown the sensitivity of $\Reff$ to single topological changes in the network near the source and sink nodes.  In this section, we show that the resistance of a network is less sensitive to individual topological changes if more of the network topology is averaged over by a single measurement.  Specifically, we show that summing $\Reff$ calculated between every pair of nodes in the network, the total effective resistance in Eq.~\eqref{eq:Rtot}, is less sensitive to single topological changes.  Similarly, increasing the network size decreases the sensitivity of $\Reff$ to single topological changes.

\begin{figure}[h]
\centering
\includegraphics[width=0.48\textwidth]{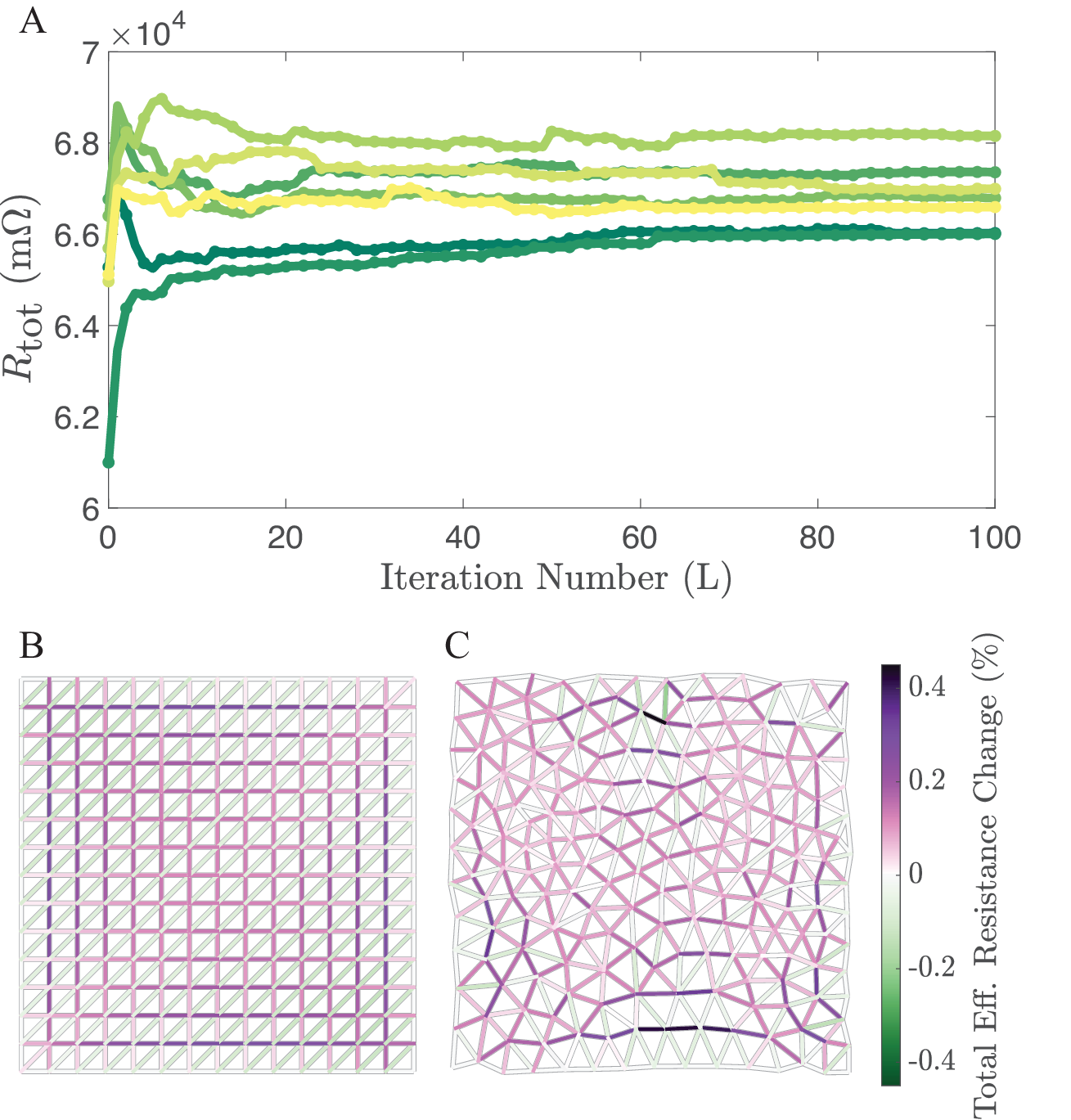}
\caption{A: Total effective resistance, $\Rtot$ computed from Eq.~\eqref{eq:Rtot}, as a function of iteration number L.  Each curve is the evolution of a different initial point cloud.  B: 
Heatmap where the color of each edge indicates the relative percent change of the total resistance, $\Rtot$, when the edge is flipped to the other diagonal of its quadrilateral.  
Same network as Fig.~\ref{fig:LatticeHeatMap}A.
C: Same as B but for a disordered network at $L=10$.
In both B and C, boundary edges and those not part of a convex quadrilateral are not flipped, and are colored white (i.e. zero). 
}
\label{fig:N200RtotEnsemble}
\end{figure}

In Fig.~\ref{fig:N200RtotEnsemble}A, we plot the total effective resistance, Eq.~\eqref{eq:Rtot}, as a function of Lloyd's iteration number $L$ with each curve representing a different initial point cloud.  Comparing to Fig.~\ref{fig:N200Ensemble}, we see overall a smaller spread of values and a noticeable lack of large jumps between sequential Lloyd's iterations.
The results of conducting the edge-flip test on both the square lattice with diagonals and a sample random network (generated with the method described in Sec.~\ref{sec:networkgeneration}) are shown in Figs.~\ref{fig:N200RtotEnsemble}B and C, respectively.
Note the decrease in percentage change from around 10\% maximum for $\Reff$ down to  0.35\% for $\Rtot$.

While the overall importance of a single edge is reduced in the case of $\Rtot$ relative to $\Reff$, we still observe how local topological changes affect this global measure of the network's electrical transport.
Figure \ref{fig:N200RtotEnsemble}B shows the result of the edge flip test for the square lattice with diagonals across each cell (same network as Fig.~\ref{fig:LatticeHeatMap}A). 
Here, the network itself has directionality and $\Rtot$, being a global quantification of the network, reflects this directionality in the response of $\Rtot$ to the edge flip test.
Diagonal edges near the northwest and southeast corners have a larger effect than other diagonal edges.
Additionally, we see that edges parallel to the boundary have a large effect, with this effect being largest for the middle of the outermost ring of interior edges.  The horizontal and vertical edges having a positive change on $\Rtot$ is consistent with these edges increasing in length, which is proportional to resistance, when flipped without creating any shorter paths through the network.  The diagonal edges having a negative change on $\Rtot$ is consistent with the addition of a new shortest paths through this alternative diagonal.
We see similar trends in the disordered network in Fig.~\ref{fig:N200RtotEnsemble}C.  
As with $\Reff$, we see that individual topological changes to the network effect $\Rtot$ in different ways depending on the edge's location and orientation, but to a smaller degree.

\begin{figure}[h!]
\centering
\includegraphics[width=0.41\textwidth]{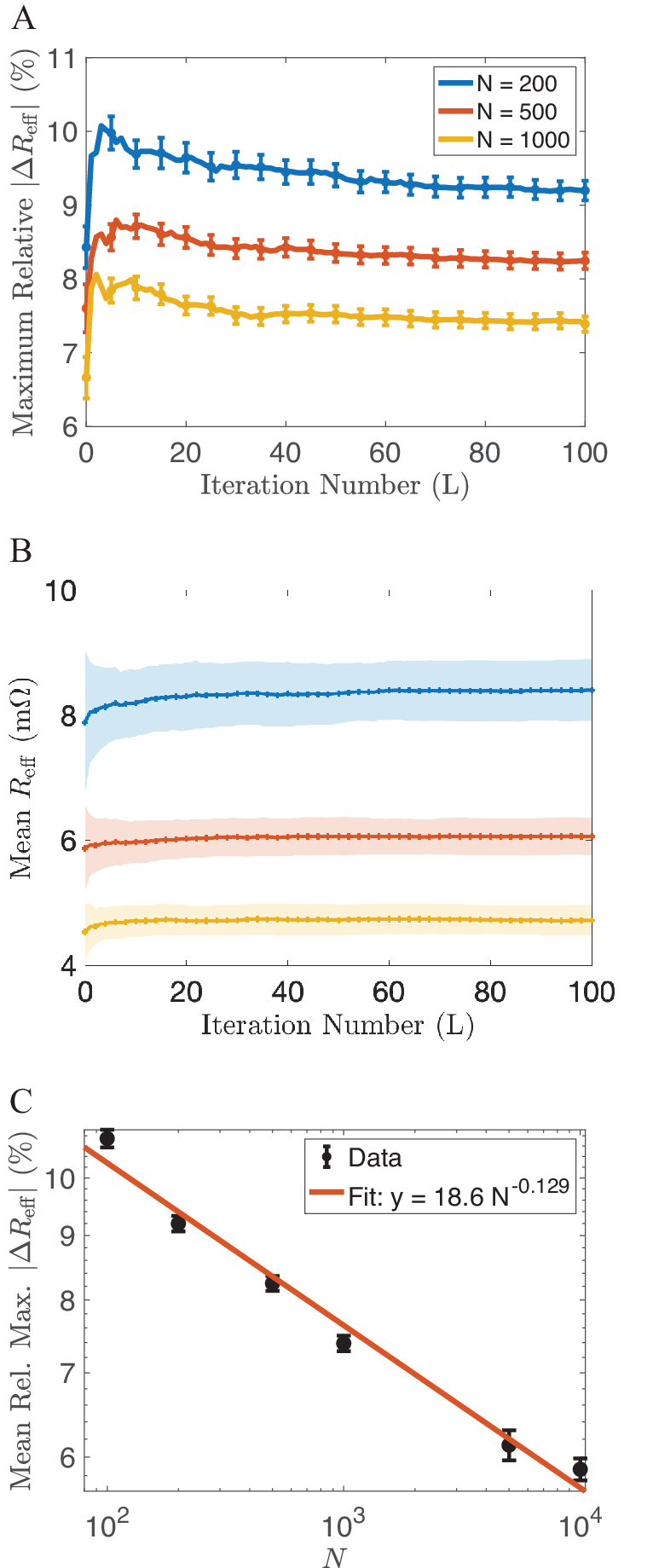}
\caption{A: Maximum effective resistance change as a function of iteration number L across an ensemble of 100 networks for $N=200$, $N=500$, and $N=1000$ under the local edge-flip test. The curves correspond to the means, and the error bars for every fifth iteration correspond to the standard errors. B: Effective resistance as a function of iteration number L for the same ensembles as panel A.  The curves correspond to the means, and the shaded regions correspond to the standard deviations. C: Mean of maximum effective resistance change at iteration number $L = 100$, as a function of point cloud size $N$, with error bars corresponding to the standard errors.  For $N=5,000$ and $N=10,000$ only 20 simulations were averaged.}
\label{fig:Finite_size}
\end{figure}

Another way to decrease the importance of a single edge is to increase the total number of edges by creating larger networks with more points inside the same-sized box.  This increase in edge number decreases the overall size of each edge and also increases the number of pathways between any two nodes, thus intuitively changing any one edge should have a smaller effect. 
This is demonstrated in Fig.~\ref{fig:Finite_size}A, which shows the result of performing the edge-flip test and recording the maximum magnitude of percent change in $\Reff$ for each of 100 networks at each Lloyd's iteration number. Given the prediction from \ref{sec:predict}, to reduce computational complexity, only quadrilaterals that contain source or sink nodes are flipped. If there does not exist a convex quadrilateral that contains source or sink nodes, the enumeration edge-flip test is conducted to record the maximum change. The curve is the average over this ensemble, and the error bar is one standard error above and below. Note that increasing the network size also decreases the effective resistance, as shown in Fig.~\ref{fig:Finite_size}B. We also notice a slight increasing of the mean value of $\Reff$ as a function of L. A similar increase in the resistance of Delaunay-based networks with increasing local translational order has been observed in Ref.~\cite{raj_local_2025} for a different type of point cloud. In Fig.~\ref{fig:Finite_size}C, we show that the mean relative maximum change in $\Reff$ decreases with $N$, like  $N^{-0.129}$, suggesting that it will vanish as $N \to \infty$. 
Understanding finite-size effects on $\Reff$ (and $\Rtot$) is important from a practical standpoint because of the size limitations imposed by the processes used to create network metamaterials, which typically involve additive manufacturing (see, e.g., \cite{obrero_electrical_2025, moody_methodology_2025, shen_autonomous_2024}).

\subsection{Voronoi Tessellation}
\label{sec:voronoi}

As described in Fig.~\ref{fig:Mechanism}, the local topological changes associated with Delaunay networks result in abrupt changes in effective resistance under point cloud perturbations.  While the effect of local topological changes on the effective and total resistance of the network is universal, we only observe such large jumps between Lloyd's algorithm iterations when constructing networks with the Delaunay triangulation.
To illustrate this, we discuss $\Reff$ of a Voronoi tessellation on the underlying point cloud.

\begin{figure}[h!]
\centering
\includegraphics[width=0.45\textwidth]{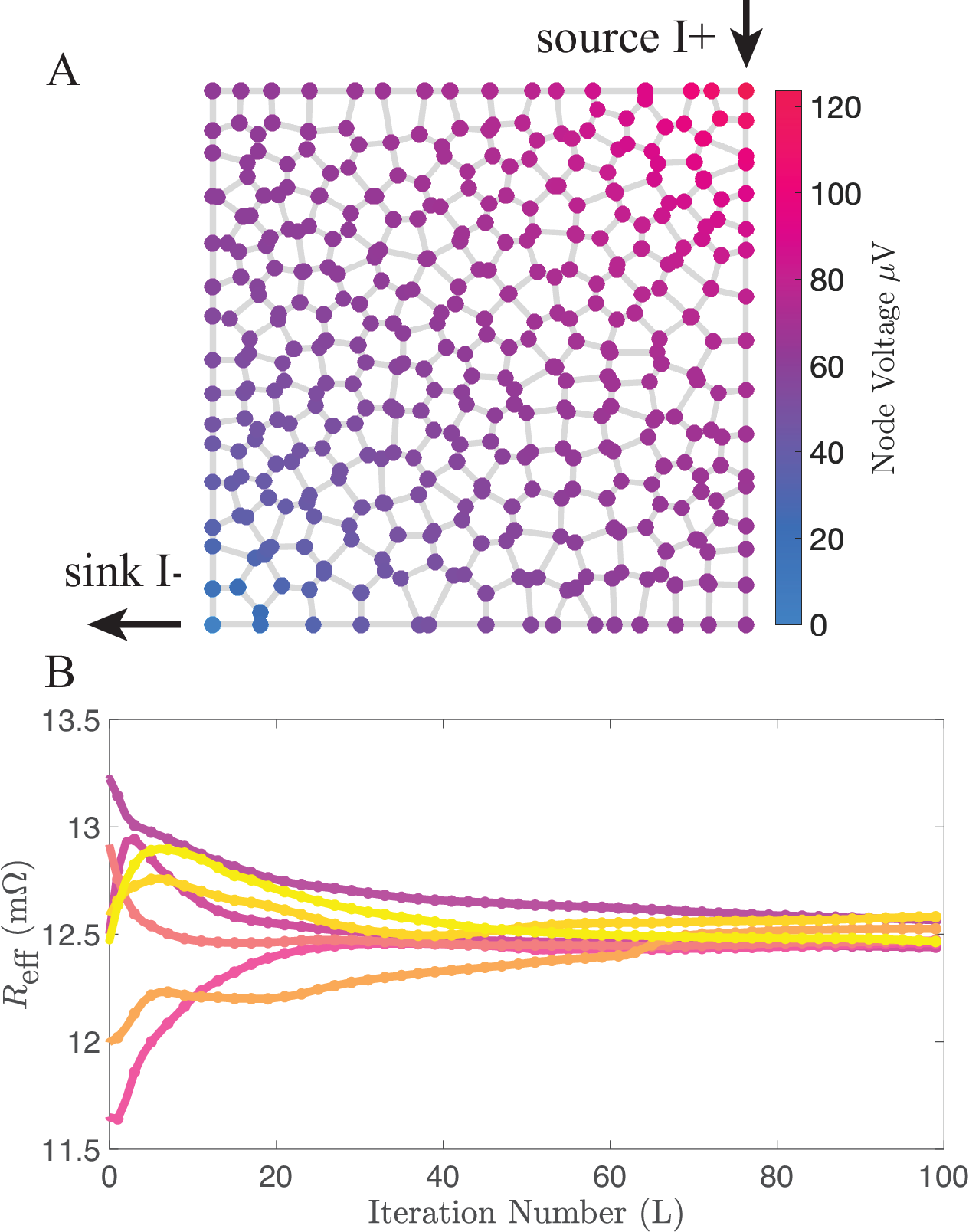}
\caption{A: Schematic showing the source and sink nodes used to compute $\Reff$ diagonally across the Voronoi network with the same point pattern as Fig.~1A. Nodes are colored by their voltage, computed with Eq.~\eqref{eq:LVI}. B: Effective resistance, $\Reff$, as a function of iteration number L for Voronoi tessellated networks. Each curve is the
evolution of a different initial point cloud. 
 These Voronoi networks are generated with the same point clouds that generated the previous Delaunay networks.}
\label{fig:Voronoi}
\end{figure}

With the same 20 simulations of a point cloud evolved under 100 iterations of Lloyd's algorithm used in Fig. \ref{fig:N200Ensemble}, we form the networks using a Voronoi tessellation constructed inside the bounded square domain. 
Figure~\ref{fig:Voronoi}A depicts the measurement of $\Reff$ across the northeast-southwest diagonal of the aforementioned networks.
In stark contrast to the Delaunay networks 
shown in Fig.~\ref{fig:N200Ensemble}B, the Voronoi networks' effective resistances in Fig.~\ref{fig:Voronoi}B evolve more smoothly as Lloyd's algorithm is iteratively applied.  Small perturbations to the point cloud in turn cause small perturbations to the length of edges in the Voronoi polygons, with topological changes (addition or removal of edges) occurring as a limit of vanishing edge length.
By contrast, the topological changes in the Delaunay triangulation tend to involve longer edges, which tends to result in more drastic changes in $\Reff$.
Additionally, we note that despite the smaller variance of $\Reff$ at large L in Fig.~\ref{fig:Voronoi} than in Fig.~\ref{fig:N200Ensemble}B, the values of $\Reff$ for individual simulations in Fig.~\ref{fig:Voronoi} should in general \textit{not} converge to a single value because Lloyd's algorithm with disordered initial conditions has been universally shown to yield disordered configurations in two dimensions 
\cite{klatt_universal_2019}.
Moreover, the magnitude of $\Reff$ in Fig.~\ref{fig:Voronoi} is larger than what is observed in Fig.~\ref{fig:N200Ensemble}B, which can be attributed to different connectivity patterns between the types of tessellations.
Since our focus in this work is on the effect of local topological changes in networks as opposed to a more comprehensive study of the tessellations themselves, we refer readers to Ref. \cite{raj_local_2025} for a more detailed discussion of this point.

\subsection{Local Tortuosity Change Correlates with Size of Effective Resistance Jumps}

\begin{figure}[h!]
\centering
\includegraphics[width=0.45\textwidth]{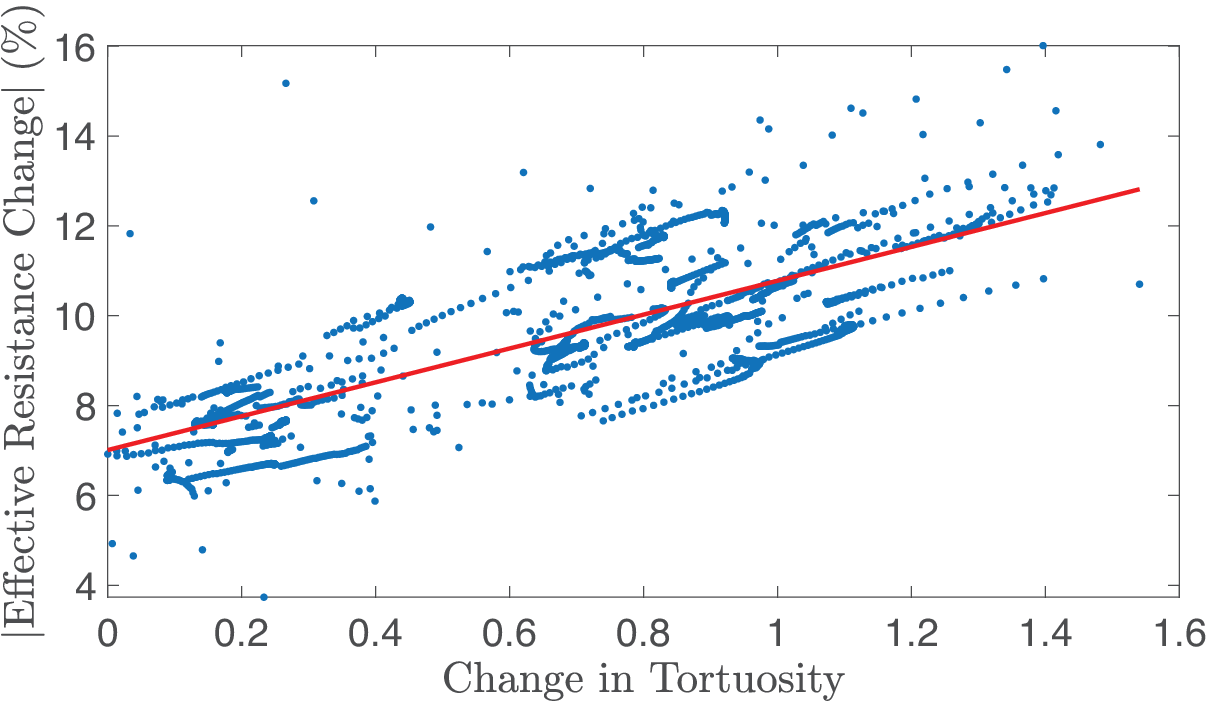}
\caption{Scatter plot of change in quadrilateral tortuosity vs.~its change in effective resistance when its diagonal edge is flipped. Quadrilaterals are taken from Fig.~\ref{fig:Quantify}A. The solid red line is a least-squares fit line to the data with $R^2 = 0.5804$ and Pearson correlation coefficient of $r = 0.7619$, indicating a strong positive linear correlation.  Points that appear to follow linear trends on either side of the red least-squares fit line are likely 
from sequential iterations of Lloyd's algorithm.
}
\label{fig:tortuosity}
\end{figure}

In two-phase heterogeneous materials, one can show that the effective resistivity of a material is related to the tortuosity of its conducting phase (see Eq. (13.226) in Ref. \cite{torquato_random_2002a}). 
The tortuosity has been computed for disordered network-based two-phase heterogeneous materials \cite{torquato_multifunctional_2018}.
To connect with the materials literature, we examine the relationship between the changes in network tortuosity (see Eq.~(\ref{eq:delta_tau})) and effective resistance in our networks.
Having identified that network edge flips near either the source or sink nodes contribute the most to jumps in the effective resistance, we compute the tortuosity locally as opposed to over a larger portion of the network.
In Fig.~\ref{fig:tortuosity} we show the correlation between the change in tortuosity computed using Eq.~\eqref{eq:delta_tau} and the magnitude of change in effective resistance when an edge flip is performed on a quadrilateral structure in the network with one of its nodes being either the source or the sink node.  
The strong linear correlation ($r = 0.7619$ and $R^2=0.5804$) suggests that the local geometry and shortest paths in the vicinity of the source and sink nodes are both related to large changes in the network's effective resistance.
Thus, the relationship between tortuosity and effective resistance observed for materials is also relevant to spatially embedded resistor networks.

\section{Discussion and Conclusions}\label{sec:conclusions}

In this paper, we showed how the location and directionality of a topological change in a network influence the change in the network's effective and total resistances, $\Reff$ and $\Rtot$, respectively.
We focused on Delaunay triangulation networks and the Delaunay flip topological change.  The Delaunay flip is an abrupt swapping of which diagonal in a convex quadrilateral structure in the network is connected by an edge.
While this choice was motivated by the desire to understand sharp changes in $\Reff$ observed in Ref.~\cite{obrero_electrical_2025} as the point cloud underlying the Delaunay triangulation-based network evolved under Lloyd's algorithm, isolated edge removals and edge additions apply more broadly to any network regardless of how the network was created and whether or not it is spatially embedded. 
We demonstrated this with two cases, one in which the source node was moved away from the boundary, and one in which the underlying point cloud used to generate the Delaunay triangulation was formed by uniformly randomly perturbing the points of a hexagonal lattice.
Our analytical treatment of approximating sequential edge removal and edge addition as 
two independent rank-one changes to the weighted combinatorial graph Laplacian and computing its inverse using the Sherman--Morrison formula \cite{sherman_adjustment_1950} is also broadly applicable as no assumption on the network structure itself was necessary.

Through the edge flip test and analytical treatment, we showed that edges near the source and sink nodes
for the effective resistance measured diagonally across the network, as in Ref.~\cite{obrero_electrical_2025}, had the largest effect; these edges had large voltage drops in either their original or flipped orientation.
What was more surprising was how large this effect could be, either increasing or decreasing $\Reff$ up to 10\% in the case of $N=200$ node networks.  This greatly overshadowed the effects of node movement that lengthened or shortened edges, or topological changes to any other edge outside the immediate neighborhood of either the source or sink node.
To relate our mathematical network model to the heterogeneous network materials it emulates, we computed the change in tortuosity across quadrilaterals containing the source and sink nodes and compared that to the change in effective resistance.
We found a strong positive correlation between the change in tortuosity and effective resistance, which is consistent with the expectation from the materials literature \cite{torquato_random_2002a}.

This study, focused on finite-sized network representations of two-phase open lattice materials,  departs from typical treatments of random graphs that seek global quantifiers or ensemble-averaged statistics.
Overall, we saw some relationship to shortest paths in the network, but a shortest path alone cannot completely describe either the effective resistances or predict the effect of single topological changes.  For example, decreases to $\Rtot$ were consistent with the
addition of new shortest paths after a diagonal edge was flipped, but could not predict differences in the magnitude of this change.  
Similarly, other network statistics like edge betweenness that are able to predict failure processes in materials \cite{berthier_forecasting_2019} fail to predict the magnitude of the changes in $\Reff$ \cite{obrero_electrical_2025}.   
We briefly addressed ensemble-averaged statistics, confirming that each edge's importance vanishes as the number of nodes increases.  The average edge length scales with $1/\sqrt{N}$ \cite{obrero_electrical_2025} and the number of edges scales linearly with $N$, however, the rate of vanishing importance scaled neither as $1/\sqrt{N}$ nor $1/N^{1/4}$.
Mean effective resistance was found to increase slightly with local order; see e.g., Ref.~\cite{raj_local_2025} for more relationships between transport and global network characteristics as a function of disorder and tessellation type.\\

From a material design point of view, the effect of local topological changes remains an important effect to study due to the practical limitations in network size imposed by the additive manufacturing methods commonly used to make network metamaterials (e.g., in Refs. \cite{obrero_electrical_2025, moody_methodology_2025, shen_autonomous_2024}). 
Understanding which local regions of the network produce the largest change can guide optimization routines in this high-dimensional design space to manufacture disordered network metamaterials with tailored transport properties.
Typical gradient-based optimization algorithms acting on the node positions will encounter issues due to the large jumps in $\Reff$ that accompany the Delaunay triangulation topological changes.
We showed that using a Voronoi tessellation instead removes these abrupt changes and is thus less problematic for use in optimization algorithms.
If instead an optimization is desired over a network configuration with fixed nodes and movable edges, the prediction of replacing one edge with another and its proportionality to voltage differences across the old and new edge remains applicable regardless of whether or not the edges are the two diagonals of a convex quadrilateral.


\begin{acknowledgments}
The authors would like to thank Karen E. Daniels for insightful discussions and Mason A. Porter for helpful feedback on the manuscript. This work is supported by the collaborative NSF DMREF Grant No. CMMI-2323342 and NSF Grant No. DMS-2307297.
\end{acknowledgments}

\section*{Data Availability}

The code to generate and evolve a point cloud using Lloyd's algorithm is available on GitHub \cite{GitHub_config_generation}.  The code to compute the effective resistance, total effective resistance, and execute the edge flip enumeration test is available on GitHub \cite{GitHub_config_resistance}.  The set of 20 $N=200$ node networks is generated from the same evolving point clouds used in Ref.~\cite{obrero_electrical_2025} and are available on Dryad \cite{Wang_data}.


%

\end{document}